%% file: main.tex
\documentclass[sigconf,natbib=true]{acmart}

\usepackage{multirow}
\usepackage{color, colortbl}
\usepackage{arydshln}
\usepackage{array}
\usepackage{balance} 
\usepackage{subcaption}
\usepackage{pifont}
\usepackage{amsmath}
\usepackage{graphicx}
\usepackage[most]{tcolorbox}
\usepackage{xcolor}

\AtBeginDocument{%
  }

\setcopyright{acmlicensed}
\copyrightyear{2025}
\acmYear{2025}
\acmDOI{XXXXXXX.XXXXXXX}
\acmConference[IR-RAG 2025]{Information Retrieval's Role in RAG Systems}{July 17,
  2025}{Padua, Italy}

\input{_abbrev.tex}

\begin{document}

\title{Walk\&Retrieve: Simple Yet Effective Zero-shot Retrieval-Augmented Generation via Knowledge Graph Walks}

\author{Martin Böckling}
\email{martin.boeckling@uni-mannheim.de}
\orcid{0000-0002-1143-4686}
\affiliation{%
  \institution{University of Mannheim}
  \city{Mannheim}
  \country{Germany}
}

\author{Heiko Paulheim}
\email{heiko.paulheim@uni-mannheim.de}
\orcid{0000-0003-4386-8195}
\affiliation{%
  \institution{University of Mannheim}
  \city{Mannheim}
  \country{Germany}
}

\author{Andreea Iana}
\email{andreea.iana@uni-mannheim.de}
\orcid{0000-0002-7248-7503}
\affiliation{%
  \institution{University of Mannheim}
  \city{Mannheim}
  \country{Germany}
}

\renewcommand{\shortauthors}{Böckling et al.}

\begin{abstract}
    Large Language Models (\llms) have showcased impressive reasoning abilities, but often suffer from hallucinations or outdated knowledge. Knowledge Graph (\kg{})-based Retrieval-Augmented Generation (\rag{}) remedies these shortcomings by grounding \llm{} responses in structured external information from a knowledge base. However, many \kg{}-based \rag{} approaches struggle with (i) aligning \kg{} and textual representations, (ii) balancing retrieval accuracy and efficiency, and (iii) adapting to dynamically updated \kgs. 
    In this work, we introduce \walkretrieve{}, a simple yet effective \kg-based framework that leverages walk-based graph traversal and knowledge verbalization for corpus generation for zero-shot \rag{}. Built around efficient \kg{} walks, our method does not require fine-tuning on domain-specific data, enabling seamless adaptation to \kg{} updates, reducing computational overhead, and allowing integration with any off-the-shelf backbone \llm. Despite its simplicity, \walkretrieve{} performs competitively, often outperforming existing \rag{} systems in response accuracy and hallucination reduction. Moreover, it demonstrates lower query latency and robust scalability to large \kgs, highlighting the potential of lightweight retrieval strategies as strong baselines for future \rag{} research.
\end{abstract}

\ccsdesc[500]{Information systems~Information retrieval}
\ccsdesc[300]{Information systems~Language models}
\ccsdesc[300]{Information systems~Question answering}

\keywords{Knowledge Graph Retrieval-Augmented Generation, Graph Walks, Zero-Shot Retrieval, Question Answering}

\maketitle

\input{sections/introduction}
\input{sections/methodology}
\input{sections/experimental_setup}
\input{sections/results_discussion}
\input{sections/conclusion}

\bibliographystyle{ACM-Reference-Format}
\balance
\bibliography{references}

\end{document}

%% file: _abbrev.tex
\newcommand{\rparagraph}[1]{\vspace{1.2mm}\noindent\textbf{#1.}}
\newcommand{\iparagraph}[1]{\vspace{1.2mm}\noindent\textit{#1.}}

\usepackage{todonotes}

\definecolor{Gray}{gray}{0.9}
\definecolor{ashgrey}{rgb}{0.7, 0.75, 0.71}
\definecolor{darkgreen}{rgb}{0,0.39,0}

\newcolumntype{g}{>{\columncolor{Gray}}c}

\newcommand{\walkretrieve}{\texttt{Walk\&Retrieve}}
\newcommand{\wrrandom}{\texttt{Walk\&Retrieve-RW}}
\newcommand{\wrbfs}{\texttt{Walk\&Retrieve-BFS}}

\newcommand{\rra}{RetrieveRewriteAnswer}
\newcommand{\subgraphrag}{SubgraphRAG}
\newcommand{\rag}{RAG}
\newcommand{\qa}{QA}
\newcommand{\kg}{KG}
\newcommand{\kgs}{KGs}
\newcommand{\llm}{LLM}
\newcommand{\llms}{LLMs}

%% file: sections/introduction.tex
\section{Introduction}
\label{sec:introduction}
Large Language Models (\llms) are pivotal to question answering (\qa) due to their strong language understanding and text generation capabilities \cite{brown2020language,ouyang2022training,zhao2023survey,liu2023pre,touvron2023llama2,lievin2024can}. 
However, \llms{} often (i) struggle with outdated knowledge, (ii) lack interpretability due to their black-box nature \cite{danilevsky2020survey}, and (iii) can hallucinate convincingly yet factually inaccurate answers \cite{rawte2023troubling,ji2023survey,huang2024survey}. These issues are particularly pronounced in knowledge-intensive tasks \cite{mallen2023not}, when dealing with domain-specific \cite{tonmoy2024comprehensive,sun2024head} or rapidly changing knowledge \cite{vu-etal-2024-freshllms}.
Retrieval-augmented generation (\rag) mitigates these limitations by grounding responses in relevant external information \cite{lewis2020retrieval,gao2023retrieval,fan2024survey}. 
Yet, text-based \rag{} primarily relies on semantic similarity search of textual content \cite{fan2024survey}, which fails to capture the relational knowledge necessary to integrate passages with large semantic distance from the query in multi-step reasoning \cite{larson2024,chen2024benchmarking,jin2024graph,peng2024graph,ma2024think}. 

Consequently, several works leverage knowledge graphs (\kgs) -- structured knowledge bases representing real-world information as networks of entities and relations \cite{hogan2021knowledge} -- as external information sources to overcome standard \rag{} limitations \cite{peng2024graph}. 
Given a query, KG-based \rag{} systems retrieve relevant facts as nodes, triplets, paths, or subgraphs using graph search algorithms, or parametric retrievers based on graph neural networks or language models \cite{peng2024graph}. The retrieved graph data is then reformatted for the language model -- via linearized triples \cite{kim2023kg}, natural language descriptions \cite{wu2023retrieve,li2023graph,wen2023mindmap,edge2024local,fatemi2024talk}, code-like forms \cite{guo2023gpt4graph}, or node sequences \cite{luoreasoning,sunthink,mavromatis2024gnn} -- and finally used by an \llm{} to generates the final response \cite{peng2024graph}.

\begin{figure*}[t]
  \centering
  \includegraphics[width=\textwidth]{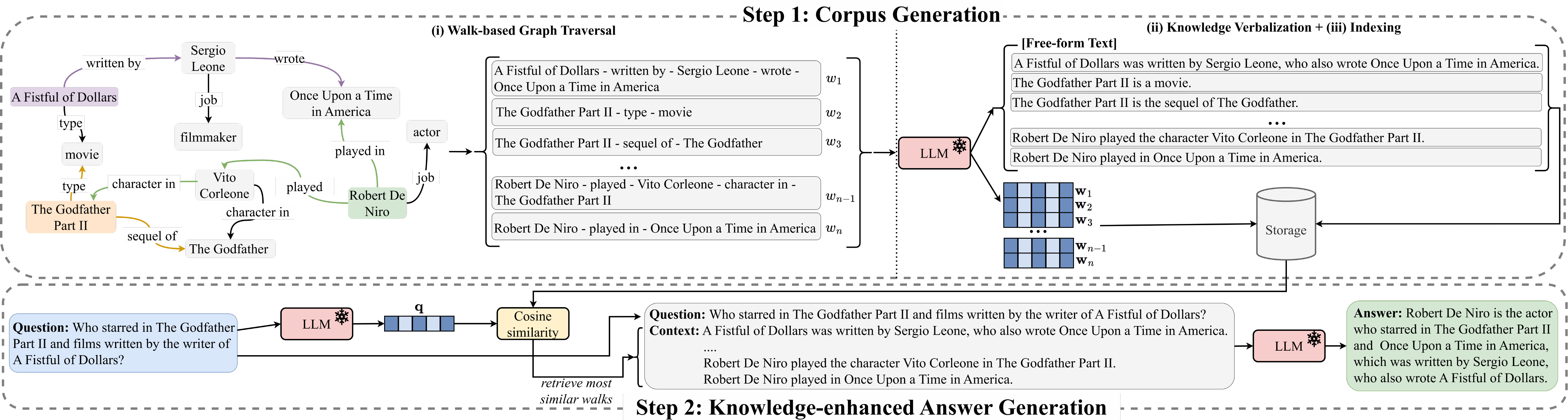}
  \caption{Overview of the \walkretrieve{} framework: (1) We combine walk-based graph traversal with knowledge verbalization for corpus generation; (2) The answer is generated with a prompt augmenting the query with the most similar verbalized walks.
  }
  \label{fig:framework}
  \vspace{-1em}
\end{figure*}

The existing body of work exhibits several drawbacks.
First, augmenting a query with relevant \kg{} triples \cite{sen2023knowledge,saleh-etal-2024-sg,li2024simple} can lead to suboptimal retrieval performance due to the misalignment of structured graphs and the sequential token-based nature of the language model. 
Although converting \kg{} data to a LLM-suitable tokenized format can help, naive triple linearization \cite{he2024g,baek2023knowledge}, which directly converts \kg{} triples into plain text without considering context, coherence, or structural nuances, often produces semantically incoherent descriptions \cite{wu2024cotkr}.\footnote{Given the triples: \texttt{A Fistful of Dollars $\rightarrow$ writtenBy $\rightarrow$ Sergio Leone}, and \texttt{The Godfather Part II $\rightarrow$ sequelOf $\rightarrow$ The Godfather}, a prompt based on naive linearization would be: \textit{These facts might be relevant to answer the question: (A Fistful of Dollars, writtenBy, Sergio Leone), (The Godfather Part II, sequelOf, The Godfather)}[...].}
Second, \rag{} systems that directly reason over \kgs{} with \llms{} perform a step-by-step graph traversal for fact retrieval \cite{jin2024graph,sunthink,ma2024think}. This requires multiple \llm{} calls per query, significantly increasing complexity and latency. 
Third, KG-based \rag{} models often fine-tune retrievers \cite{wu2023retrieve,guo2024knowledgenavigator,luoreasoning} or generators \cite{yasunaga2021qa,he2024g,hu2024grag,luoreasoning,mavromatis2024gnn} on task-specific data to better adapt to diverse \kg{} structures and vocabularies. However, collecting high-quality instruction data is costly \cite{cao2023instruction}, and fine-tuning large models -- even with parameter-efficient methods \cite{hulora,chai2023graphllm,perozzi2024let} -- is expensive and limits generalization to dynamic KGs or unseen domains \cite{li2024simple,wu2023retrieve}.

\rparagraph{Contributions}
%
%
We propose \walkretrieve{}, a lightweight zero-shot \kg{}-based \rag{} framework, designed as a simple yet competitive baseline to address these challenges.
It combines efficient graph traversal, via random or breadth-first search walks, with verbalization of \kg{}-derived information to build a contextual corpus of relevant facts for each \kg{} entity. At inference, we retrieve the most similar nodes to the query, and their corresponding walks, respectively. We generate the final answer by prompting an \llm{} with the query, augmented with this relevant context.  
Unlike many existing \kg{}-based \rag{} systems, \walkretrieve{}:
\textbf{(1)} is \textit{adaptable} to dynamic \kgs{} -- updates (e.g., node insertion or deletion) require no retraining, as new knowledge can be added by incrementally generating additional walks;
\textbf{(2)} is more \textit{efficient}, requiring no fine-tuning of the backbone \llm{}, and only a single \llm{} call per query;
\textbf{(3)} enables \textit{zero-shot RAG} with any \textit{off-the-shelf \llm{}}.
We show that \walkretrieve{} consistently generates accurate responses, while minimizing hallucinations. Our findings render walk-based corpus generation as a promising approach for scalable \kg-based \rag{}, and establish \walkretrieve{} as a strong baseline for future research.

%% file: sections/methodology.tex
\section{Methodology}
\label{sec:methodology}
Fig. \ref{fig:framework} illustrates our proposed framework, comprising two stages: corpus generation and knowledge-enhanced answer generation.  

\subsection{Corpus Generation}
\label{sec:corpus_generation}
In the first stage, we leverage the knowledge stored in \kgs{} to construct a corpus of relevant facts. A Knowledge Graph is defined as $G = (V, E, R)$, where $V$ denotes a set of nodes $v \in V$, and $E \subseteq V \times R \times V$ a set of directed edges labeled with relation types from the set $R$. For each node $v \in V$, we define its neighbor set as $N(v) := \{v': \exists r \in R | (v, r, v') \in E\}$.
Corpus generation consists of walk-based graph traversal, knowledge verbalization, and indexing.

\rparagraph{Walk-based Graph Traversal}
We extract relevant facts for all entities in the \kg{} using two walk-based graph traversal approaches. 

\iparagraph{Random Walks (RW)} 
In this method, we retrieve facts for a given vertex $v \in V$ by generating $n_w$ graph walks $\mathcal{W}_l$ of length $l$ rooted in $v$. 
A random walk is a stochastic process with variables $X_0, X_1, X_2, ..$, where each $X_t \in V$ denotes the vertex visited at time $t$ \cite{perozzi2014deepwalk}.
At each step, when the random walker is at vertex $v_i$, it chooses the next node uniformly at random from one of its neighbors $v_j \in N(v_i)$ according to the following transition probability: 
\vspace{-0.3em}
\begin{equation}
    P(X_{t+1} = j | X_t = i) = 
        \begin{cases}
            \frac{1}{|N(v_i)|} \text{ if} (v_i, r, v_j) \in E \\
            0 \text{ otherwise},
        \end{cases}
\end{equation}
where $|N(v_i)|$ denotes the neighborhood size of $v_i$.
Finally, the graph corpus is obtained by aggregating $n_w$ random walks $\mathcal{W}_l = (X_0, r_i, X_1, ..., r_k, X_l), r \in R$ per vertex, as $\mathcal{C}_{RW} = \bigcup_{i=1}^{|V|} \bigcup_{j=1}^{n_{walks}} \mathcal{W}_l$. 

\iparagraph{Breadth-First Search (BFS) Walks}
In this approach, we construct a spanning tree for each entity in $G$ using the BFS algorithm. 
For a given root $v_r \in V$, we build walks by partitioning the reachable nodes $v_j$ into layers $L_i$ based on their shortest-path distance to the root \cite{ristoski2016rdf2vec}. 
Starting with $L_0 = \{v_r \}$, layers are recursively defined as
\begin{equation}
    L_{i+1} = \{v_j \in V \backslash \cup_{k=0}^i L_k: \exists v_i \in L_i | (v_i, r, v_j) \in E\}
\end{equation}
for $i \in [0, d]$, where $d$ is the maximum depth (i.e., the maximum allowed shortest-path distance).
This guarantees that each vertex is explored only once per search. 
Hence, the resulting corpus $\mathcal{C}_{BFS} = \bigcup_{i=1}^{|V|} L_i$ contains only non-duplicate walks for each vertex.
We note that the maximum allowed shortest-distance path $d$ of the BFS walks is equivalent to the length $l$ of the randomly generated walks.

\begin{figure*}[t]
  \centering

  \begin{subfigure}[t]{\textwidth}
    \centering
    \begin{tcolorbox}[
        colback=gray!20,       
        colframe=black,         
        boxrule=1pt,            
        width=\textwidth,       
        fontupper=\footnotesize,        
        boxsep=2pt,               
        left=3pt, right=3pt,      
        top=2pt, bottom=2pt       
      ]
      \textbf{System}: Please provide me from an extracted triple set of a Knowledge Graph a sentence. The triple set consists of one extracted random walk. Therefore, a logical order of the shown triples is present. Please consider this fact when constructing the sentence. Prevent introduction words. \\
      \textbf{Human}: Please return only the constructed sentence from the following set of node and edge labels extracted from the Knowledge Graph: \textcolor{blue}{\{triples\}}.
    \end{tcolorbox}
    \vspace{-1em}
    \caption{Knowledge verbalization.}
    \label{fig:prompt_template_verbalization}
  \end{subfigure}\vfill

  \begin{subfigure}[t]{\textwidth}
    \centering
    \begin{tcolorbox}[
        colback=gray!20,
        colframe=black,
        boxrule=1pt,
        width=\textwidth,
        fontupper=\footnotesize,
        boxsep=2pt,              
        left=3pt, right=3pt,      
        top=2pt, bottom=2pt      
      ]
      \textbf{System}: You are provided with context information from a RAG retrieval, which gives you the top k context information. Please use the provided context information to answer the question. If you are not able to answer the question based on the context information, please return the following sentence: ``I do not know the answer". \\
      \textbf{Human}: Please answer the following question: \textcolor{blue}{\{question\}}. Use the following context information to answer the question: \textcolor{blue}{\{context\}}.
    \end{tcolorbox}
    \vspace{-1em}
    \caption{Knowledge-enhanced answer generation.}
    \label{fig:prompt_template_answer_gen}
    
  \end{subfigure}
  \caption{Prompt templates used for knowledge verbalization and answer generation.}
  \label{fig:prompt_templates}
  \vspace{-1em}
\end{figure*}

\rparagraph{Knowledge Verbalization}
As \llms{} require textual inputs, we convert the extracted walks for each entity in $G$ into free-form textual descriptions, to enable knowledge-enhanced reasoning for answer generation.
In contrast to recent works that fine-tune an \llm{} on question-answer pairs to learn a graph-to-text transformation \cite{wu2023retrieve}, we directly prompt the \llm{} -- using the prompt template shown in Fig. \ref{fig:prompt_template_verbalization} -- to provide a natural language representation of the walks, obtaining the verbalized corpus. 
This approach aligns the \kg{}-derived information with the \llm{}'s representation space, while preserving the order of the nodes and edges in the walks. Moreover, by not fine-tuning the \llm{}, we (i) eliminate the need for labeled graph-text pairs, (ii) improve generalization to unseen \kgs{}, and (iii) enable the usage of any \llm{} in the knowledge verbalization step.

\rparagraph{Indexing}
Lastly, we index the graph for efficient retrieval. 
After knowledge verbalization, each walk $w_i^v$ of vertex $v$ is converted into a vector $\mathbf{w}^v_i$.
Moreover, we compute each node's global representation from the concatenation of its respective walks.
We store the embeddings of all nodes and corresponding walks to facilitate efficient retrieval during inference.
Crucially, our walk-based corpus generation renders \walkretrieve{} highly adaptable to dynamic \kgs{}: updates (deletions, modifications, or additions of nodes and edges) require recomputing only the walks involving the changed graph elements -- a much smaller subset than the entire corpus.

\subsection{Knowledge-enhanced Answer Generation}
\label{sec:answer_generation}
Given a query $q$, we encode it with the same \llm{} used for knowledge verbalization, so that the query and the retrieved facts share the same vector space. 
We then perform a $k$-nearest neighbor search to retrieve the $k$ most similar nodes in $G$ to $q$ and, for each node, the $k$ most relevant verbalized walks. Concretely, we define the sets of relevant nodes $V_k$ and corresponding walks $W_k$ based on the cosine similarity between the embeddings of the query $\mathbf{q}$ and each node $\mathbf{v}$ or walk $\mathbf{w}^v$, respectively. To this end, we compute:
\vspace{-0.2em}
\begin{equation}
    \begin{split}
    V_k & = argtopk_{v \in V} cos(\mathbf{q}, \mathbf{v}) \\
    W_k & = \bigcup_{v_k \in V_k} argtopk_{w^{v_k} \in \mathcal{C}} cos(\mathbf{q}, \mathbf{w}^{v_k}), 
    \end{split}
\end{equation}
\vspace{-0.3em}
where $\mathcal{C} \in \{\mathcal{C}_{RW}, \mathcal{C}_{BFS}\}$, and the \textit{argtopk} operation retrieves the $k$ nodes with the highest cosine similarity to the query.
For zero-shot inference, we design a prompt that integrates the query $q$ with the relevant context $W_k$, cf. template from Fig. \ref{fig:prompt_template_answer_gen}. Importantly, we instruct the \llm{} to refrain from responding if the context is insufficient, thereby grounding responses in the extracted structured knowledge, and reducing hallucinations. 
Finally, the prompt is fed into the previously used \llm{} to generate a response.
By avoiding \llm{} fine-tuning, we reduce computational costs and eliminate the need for task-specific training data.
Moreover, we reduce inference latency as \walkretrieve{} uses a single call to the \llm{} per query.\footnote{Note that the preprocessing step's computational overhead is a \textit{one-time} cost, as subsequent graph changes require only incremental, inexpensive updates to the corpus.}

%% file: sections/experimental_setup.tex
\section{Experimental Setup}
\label{sec:experimetal_setup}
\rparagraph{Baselines}
We compare \walkretrieve{} against three kinds of baselines: standard \llm{}, text-based \rag{}, and \kg{}-based \rag{}. 
With \textit{\llm{} only}, we test whether the \llm{} can answer questions without external data.
For \textit{Vanilla \rag{}}, following \cite{sen2023knowledge}, we uniformly sample 5 triples from all 1-hop facts of the question entities.
We consider two \kg{}-\rag{} models. 
\textit{\subgraphrag{}} \cite{li2024simple} retrieves subgraphs using a MLP and parallel-triple scoring; the \llm{} then reasons over the linearized triples of the subgraph to generate a response. 
\textit{\rra{}} \cite{wu2023retrieve} uses constrained path search and relation path prediction for subgraph retrieval, which it then converts into free-form text to augment the prompt for response generation.

\rparagraph{Data}
We conduct experiments on MetaQA \cite{zhang2018variational} and CRAG \cite{yang2024crag}. 
MetaQA \cite{zhang2018variational} is a knowledge base \qa{} benchmark, with over 400K questions (single- and multi-hop), and a \kg{} containing 43K entities and 9 relation types. We use all its 1-hop, 2-hop, and 3-hop subsets with the "vanilla" question version.
CRAG \cite{yang2024crag} is a factual \qa{} benchmark for \rag{}, featuring over 4.4K question-answer pairs across five domains and eight question categories. It provides mock \kgs{} with 2.6 million entries.\footnote{In our experiments, we use the public test set of CRAG.}   
Table \ref{tab:mind_dataset} summarizes their statistics.
\input{tables/datasets}

\rparagraph{Evaluation Metrics}
We follow prior work \cite{saxena2022sequence,sen2022mintaka,li2024simple} and use Hits@1 to measure if a response includes at least one correct entity.
Additionally, we adopt the model-based evaluation setup of \citet{yang2024crag} to assess the quality of the generated answers using a three-way scoring system: \textit{accurate} (1), \textit{incorrect} (-1), or \textit{missing} (0). Exact matches are labeled \textit{accurate}; all others are evaluated with two \llms{}, \texttt{gpt-4-0125-preview} \cite{openaichagpt} and \texttt{Llama-3.1-70B-instruct} \cite{llama3modelcard}, to mitigate self-preference \cite{panickssery2024llm}. We report averages of \textit{accurate}, \textit{hallucinated}, and \textit{missing} responses, and the overall \textit{truthfulness} (i.e., accuracy minus hallucination) from the \llm{} evaluators.

\rparagraph{Implementation Details}
We retrieve $k=3$ similar nodes and walks, respectively, for answer generation.\footnote{In preliminary experiments with $k \in [1, 5]$, we found $k=3$ to be the optimal value that balances accuracy and hallucination.}
%
Our main experiments use \texttt{Llama-3.1-70B-instruct} \cite{llama3modelcard} with temperature $t=0$ and speculative decoding for all models. We perform 60 walks for random-walk corpus generation. For both \walkretrieve{} model variants, we use walks of depth 4 on MetaQA and 3 on CRAG. 
We train and evaluate the baselines using their official implementations, and conduct all experiments on two NVIDIA A6000 48 GB GPUs.\footnote{Code available at \href{https://github.com/MartinBoeckling/KGRag}{https://github.com/MartinBoeckling/KGRag}}

%% file: tables/datasets.tex
\begin{table}[t]
  \caption{Statistics of MetaQA \cite{zhang2018variational} and CRAG \cite{yang2024crag} test sets.}
  \label{tab:mind_dataset}
  \resizebox{0.8\columnwidth}{!}{%
  \begin{tabular}{lrrr|r}
    \toprule
     & \multicolumn{3}{c|}{\textbf{MetaQA}} & \multirow{2}{*}{\textbf{CRAG}} \\ 
     \cmidrule(lr){2-4} 
     
    & 1-hop & 2-hop & 3-hop & \\ 
    \midrule
    
    \# Question types
    & 13
    & 21
    & 15
    & 8
    \\
  
    \# Questions
    & 9,947
    & 14,872
    & 14,274
    & 1,335
    \\

  \bottomrule
\end{tabular}%
}
\vspace{-1em}
\end{table}

%% file: sections/results_discussion.tex
\input{tables/results}

\section{Results and Discussion}
\label{sec:results_discussion}
\begin{figure}[t]
  \centering
  \includegraphics[width=\columnwidth]{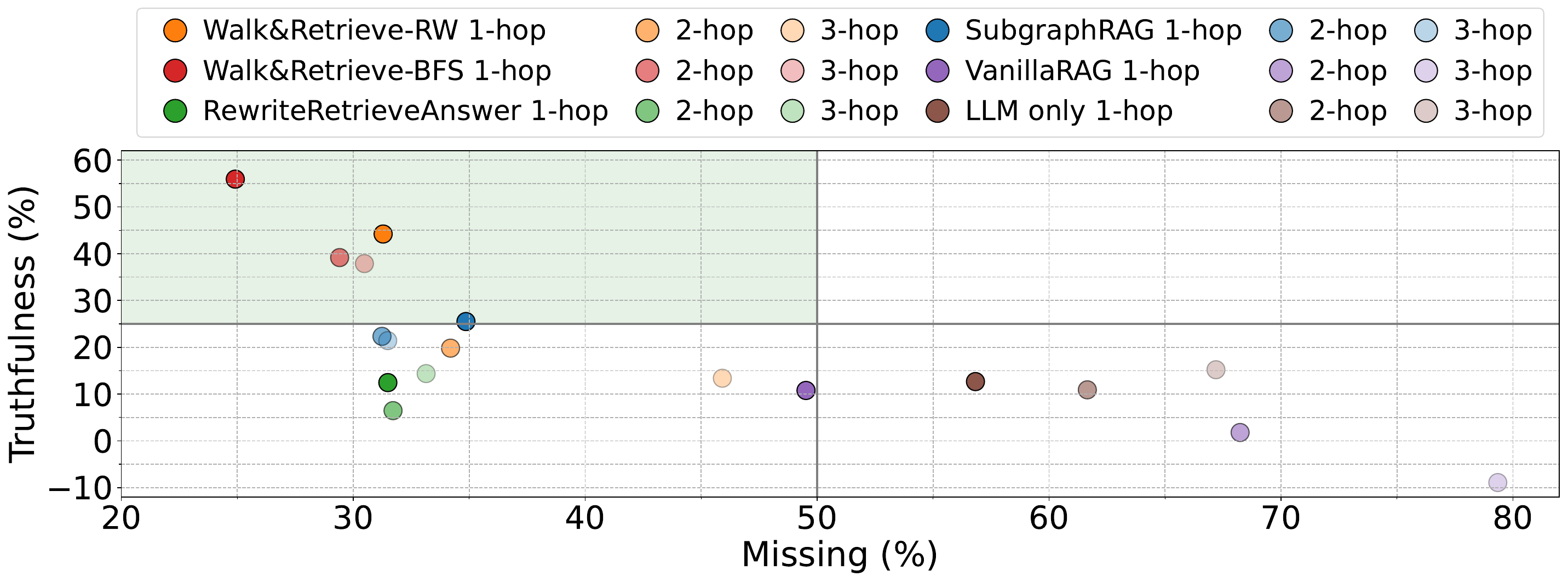}
  \caption{Missing vs. truthfulness rates over MetaQA subsets.}
  \label{fig:metaqa_performance}
  \vspace{-1.5em}
\end{figure}

Table \ref{tab:mind_dataset} summarizes the QA performance of \walkretrieve{} and the baselines with \texttt{Llama-3.1}.
On MetaQA, \wrbfs{} consistently outperforms all other models in answer accuracy and Hits@1, achieving a relative improvement of 38.64\% over the best baseline (\subgraphrag). While other \kg-based \rag{} systems yield high accuracy, they tend to hallucinate more than the simpler \llm-only and Vanilla RAG systems, which often produce no answer rather than an incorrect one. In contrast, \wrbfs{} minimizes both hallucinations and missing responses.
Although \llm-only has the lowest query latency due to the absence of a retrieval step, \walkretrieve{} achieves the fastest inference time per query among all \rag{} approaches, underscoring its efficiency.
Fig. \ref{fig:metaqa_performance} breaks down MetaQA performance by number of hops. \llm-only and Vanilla \rag{} fail to answer over 60\% of 2- and 3-hop questions. Both \subgraphrag{} and \rra{} lower the missing rate below 35\% across hops, although truthfulness remains under 25\%. Conversely, \wrbfs{} better trades off accuracy and hallucination (55\%+ truthfulness for 1-hop and 37\%+ for 2- and 3-hop questions), while greatly reducing non-responses.

On CRAG, both \walkretrieve{} variants outperform \llm-only and Vanilla \rag{} in answer accuracy, while matching them in hallucination and missing rates. Note that, \subgraphrag{} and \rra{} could not be evaluated on CRAG due to scalability and computational constraints.\footnote{\subgraphrag{} fails to scale to CRAG’s \kg{} (over 1 million edges), and \rra{} requires fine-tuning the backbone \llm{} beyond our available resources.}
%
These results highlight the scalability of our walk-based corpus generation approach, which limits traversal to small-hop neighborhoods rather than the full graph.
%
While performance drops on CRAG, likely due to its greater complexity (i.e., MetaQA expects only entity answers) and focus on holistic \rag{} performance, \walkretrieve{} remains robust. Even though the findings are promising, we plan to further evaluate \walkretrieve{} on larger \kgs{} and other challenging benchmarks (e.g., WebQSP \cite{yih2016value}, CWQ \cite{talmor2018web}) to fully showcase its capabilities.

\begin{figure}[t]
  \centering
  \includegraphics[width=\columnwidth]{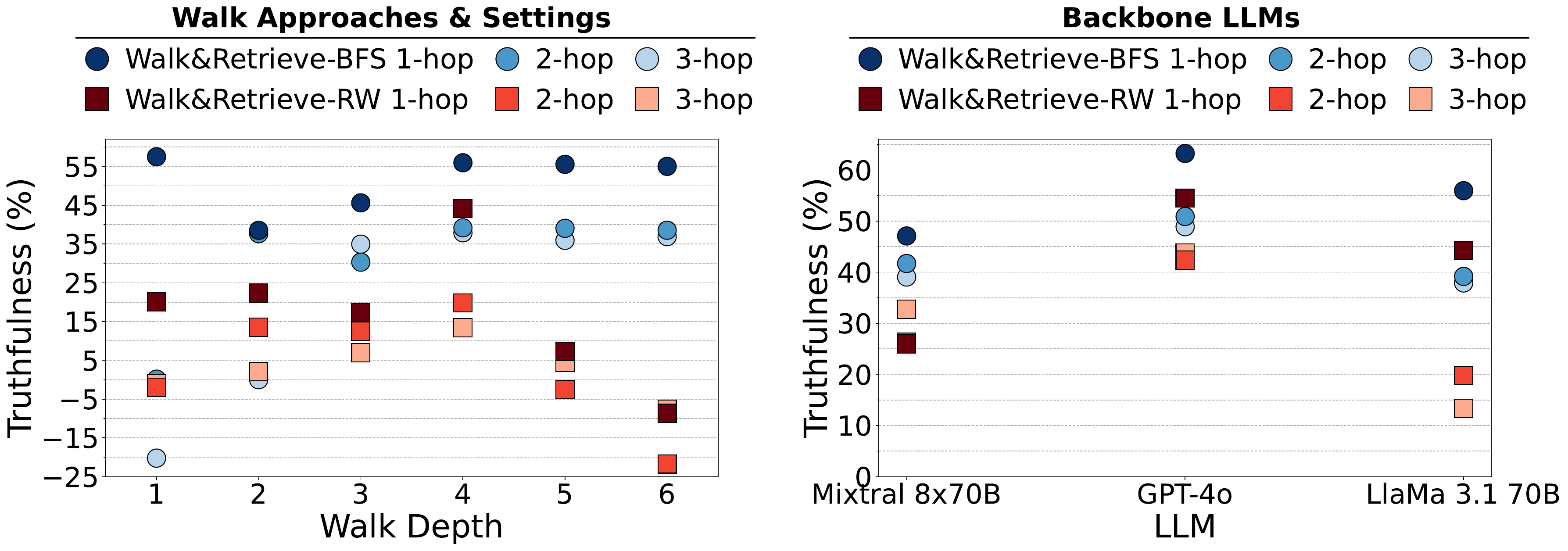}
  \caption{Truthfulness rates for different (i) walk approaches and (ii) backbone \llms{}, over the MetaQA subsets.}
  \label{fig:ablation_walk_llm}
  \vspace{-1.5em}
\end{figure}

\rparagraph{Ablation of Walk Approach}
The graph traversal strategy and its hyperparameters define a node's relevant context, directly impacting corpus quality and, consequently, retrieval accuracy in \rag{} systems.
The left graph in Fig. \ref{fig:ablation_walk_llm} shows MetaQA results for \walkretrieve{} with walk depths ranging from 1 to 6.\footnote{For \wrrandom{}, we also ablate $n_w \in [10, 100]$ (step of 10); for brevity, we report results for $n_w=60$, as other values perform comparably.}
We find that a walk depth of 4 offers the best trade-off between answer accuracy and hallucination. Notably, regardless of walk length, \wrbfs{} consistently yields higher truthfulness than \wrrandom{}, likely due to is systematic graph exploration, which avoids duplicate walks (cf. \S\ref{sec:methodology}).
In contrast, random walks tend to produce noisier context and fewer unique paths, thus capturing less relevant information from the \kg{}.\footnote{On average, each node yields 60 duplicated and 8.74 unique random walks, whereas BFS generates 9.41 unique walks. Although RW could be modified to avoid duplicates, our current setup spans the full spectrum from randomness (RW) to structure (BFS).}
While they may be more efficient on large-scale \kgs{}, as they do not compute full neighborhoods, this efficiency comes at the cost of increased noise.\footnote{The time complexity of BFS is $\mathcal{O}(|V|+|E|)$, whereas that of RW varies between $\mathcal{O}(|V|\log |V|)$ and $\mathcal{O}(|V|^{3})$.}  

\rparagraph{Robustness to Backbone \llms}
Lastly, we evaluate model robustness using different \llms{} (see right graph of Fig. \ref{fig:ablation_walk_llm}), including \texttt{Mixtral-8x7B-Instruct} \cite{jiang2023mistral} and \texttt{GPT-4o} \cite{openai2023gpt}. \texttt{Mixtral} improves answer truthfulness over \texttt{Llama-3.1} on 2- and 3-hop questions, while \texttt{GPT-4o} yields the highest truthfulness across all types of questions. The RW approach exhibits considerably higher variance across \llms{} compared to the BFS-based model, which we attribute to the noisier and less relevant information in its generated corpus. 

%% file: tables/results.tex
\begin{table*}[t]
\centering

\caption{
Question-answering performance. We report numbers in percentage, and the query runtime in seconds. For MetaQA, we average results over its k-hop subsets. The best results per column are highlighted in bold, the second best underlined.
}
\label{tab:reults}

\resizebox{\textwidth}{!}{%
    \begin{tabular}{ll|cgcgc|cgcgc}
        \toprule
        & 
        &
        \multicolumn{5}{c}{\textbf{MetaQA}} 
        & \multicolumn{5}{c}{\textbf{CRAG}} \\  
        \cmidrule(lr){3-7} \cmidrule(lr){8-12}
        
        \textbf{Baseline Type} 
        & \textbf{Model}
        & Hits@1 $\uparrow$ & Accuracy $\uparrow$ & Hallucination $\downarrow$ & Missing $\downarrow$ & Time (s) $\downarrow$
        & Hits@1 $\uparrow$ & Accuracy $\uparrow$ & Hallucination $\downarrow$ & Missing $\downarrow$ & Time (s) $\downarrow$
        \\
        \hline

        \llm{} only
        & Direct
        & 30.37  
        & 31.79  
        & 18.86  
        & 61.89  
        & \textbf{13.03}  

        & 11.05  
        & 9.31  
        & 23.95  
        & 67.49  
        & \textbf{14.14}  
        \\
        \hdashline

        Text-based \rag{}
        & Vanilla \rag
        & 25.08  
        & 14.73  
        & \underline{13.52}  
        & 65.70  
        & 22.11  

        & 15.21  
        & 16.94  
        & \textbf{19.53}  
        & \textbf{51.39}  
        & 26.01  
        \\
        \hdashline

        \multirow{4}{*}{\kg{}-based \rag{}}
        & \subgraphrag
        & 43.88  
        & \underline{41.17}  
        & 18.08  
        & 32.53  
        & 23.12  

        & -- 
        & --  
        & --  
        & --  
        & --  
        \\

        & \rra
        & 47.49  
        & 34.01  
        & 22.92  
        & \underline{32.12}  
        & 22.37  

        & -- 
        & --  
        & --  
        & --  
        & --  
        \\

        & \wrrandom{}
        & \underline{55.60}  
        & 41.11  
        & 15.31  
        & 37.13  
        & 22.12  

        & \underline{19.31} 
        & \underline{19.40}  
        & \underline{19.64}  
        & \underline{51.94}  
        & \underline{22.15}  
        \\

        & \wrbfs{}
        & \textbf{67.99}  
        & \textbf{57.08}  
        & \textbf{12.74}  
        & \textbf{28.27}  
        & \underline{21.31}  

        & \textbf{21.31}  
        & \textbf{21.53}  
        & 23.01  
        & 53.40  
        & 23.34  
        \\
        \bottomrule
        
    \end{tabular}%
    }
    \vspace{-1em}
\end{table*}

%% file: sections/conclusion.tex
\section{Conclusion}
\label{sec:conclusion}
Current \kg-based \rag{} faces challenges in aligning structured and textual representations, balancing accuracy with efficiency, and adapting to dynamic \kgs{}.
We proposed \walkretrieve{}, a simple yet effective \kg{}-based framework for zero-shot \rag. It leverages walk-based graph traversal and \llm{}-driven knowledge verbalization for corpus generation. At inference time, the \llm{} is prompted with the query augmented by relevant verbalized walks for enhanced reasoning.
Its efficient retrieval mechanism supports seamless adaptation to evolving \kgs{} through incremental generation of new walks. \walkretrieve{} is compatible with any off-the-shelf \llm{}, and reduces computational overhead by avoiding fine-tuning of the backbone \llm{}.
Despite its simplicity, \walkretrieve{} outperforms existing \rag{} approaches in answer accuracy and in the reduction of hallucinated or missing responses, while maintaining low query latency. Our results highlight walk-based corpus generation as a promising strategy for scaling to large-size \kgs{}. 
These findings establish \walkretrieve{} as a simple, yet strong baseline for \kg-based \rag{}, and we hope they inspire further research into adaptable and scalable \rag{} systems.